\documentclass[onecolumn,superscriptaddress]{revtex4-1}
\usepackage{textcomp}
\usepackage{soul}
\usepackage[english]{babel}
\usepackage[font=small]{caption}
\usepackage{amsmath}
\usepackage{graphicx}
\begin{document}

\title{Injection locking of multiple auto-oscillation modes in a tapered nanowire spin Hall oscillator}

\author{Kai Wagner}
\affiliation{Helmholtz-Zentrum Dresden - Rossendorf, Institute of Ion Beam Physics and Materials Research, 01328 Dresden, Germany}
\affiliation{TU Dresden, 01328 Dresden, Germany}
\author{Andrew Smith}
\affiliation{Department of Physics and Astronomy, University of California, Irvine, CA 92697, USA}
\author{Toni Hache}
\affiliation{Institut f\"ur Physik, Technische Universit\"at Chemnitz, D-09107 Chemnitz}
\affiliation{Helmholtz-Zentrum Dresden - Rossendorf, Institute of Ion Beam Physics and Materials Research, 01328 Dresden, Germany}
\author{Jen-Ru Chen}
\affiliation{Department of Physics and Astronomy, University of California, Irvine, CA 92697, USA}
\author{Liu Yang}
\affiliation{Department of Physics and Astronomy, University of California, Irvine, CA 92697, USA}
\author{Eric Montoya}
\affiliation{Department of Physics and Astronomy, University of California, Irvine, CA 92697, USA}
\author{Katrin Schultheiss}
\affiliation{Helmholtz-Zentrum Dresden - Rossendorf, Institute of Ion Beam Physics and Materials Research, 01328 Dresden, Germany}
\author{J\"urgen Lindner}
\affiliation{Helmholtz-Zentrum Dresden - Rossendorf, Institute of Ion Beam Physics and Materials Research, 01328 Dresden, Germany}
\author{J\"urgen Fassbender}
\affiliation{TU Dresden, 01328 Dresden, Germany}
\affiliation{Helmholtz-Zentrum Dresden - Rossendorf, Institute of Ion Beam Physics and Materials Research, 01328 Dresden, Germany}
\author{Ilya Krivorotov}
\affiliation{Department of Physics and Astronomy, University of California, Irvine, CA 92697, USA}
\author{Helmut Schultheiss}
\affiliation{TU Dresden, 01328 Dresden, Germany}
\affiliation{Helmholtz-Zentrum Dresden - Rossendorf, Institute of Ion Beam Physics and Materials Research, 01328 Dresden, Germany}



\begin{abstract}
Spin Hall oscillators (SHO) are promising candidates for the generation, detection and amplification of high frequency signals, that are tunable through a wide range of operating frequencies. They offer to be read out electrically, magnetically and optically in combination with a simple bilayer design. Here, we experimentally study the spatial dependence and spectral properties of auto-oscillations in SHO devices based on Pt(7\,nm)/\ Ni$_{\mathrm{80}}$Fe$_{\mathrm{20}}$(5\,nm) tapered nanowires.
Using Brillouin light scattering microscopy, we observe two individual self-localized spin-wave bullets that oscillate at two distinct frequencies (5.2 GHz and 5.45 GHz) and are localized at different positions separated by about 750 nm within the SHO. This state of a tapered SHO has been predicted by a Ginzburg-Landau auto-oscillator model
 , but not yet been directly confirmed experimentally. We demonstrate that the observed bullets can be individually synchronized to external microwave signals, leading to a frequency entrainment, linewidth reduction and increase in oscillation amplitude for the bullet that is selected by the microwave frequency. At the same time, the amplitude of other parasitic modes decreases, which promotes the single-mode operation of the SHO. Finally, the synchronization of the spin-wave bullets is studied as a function of the microwave power. 
We believe that our findings promote the realization of extended spin Hall oscillators accomodating several distinct spin-wave bullets, that jointly cover an extended range of tunability.


\end{abstract}

\maketitle
%
%
\thispagestyle{empty}

\section*{Introduction}

Spin current-driven magnetization dynamics has a great potential for advancing nano-sized magnetic devices towards practical applications \cite{Locatelli:2014cn,Miwa:2014hl,Choi:2014ed,Urazhdin:2014ff,Slavin:2009hb,Liu:2015ip}. As a prominent example, spin Hall oscillators (SHO) could serve as the next generation of oscillators that can be read out electrically, optically and magnetically. They unite a simple device design based on a bilayer of a heavy metal and ferromagnet with exceptionally high tuneability. Not only can this be utilized to generate versatile, yet controllable, electrical microwave signals, but is also promising for the realization of highly flexible spin-wave sources \cite{Urazhdin:2016gv}.

The active element of a SHO is a ferromagnetic layer, whose magnetization is driven by a spin current into a steady-state precession. The spin current is generated from a charge current in the adjacent heavy metal via the spin Hall effect \cite{Hoffmann:2013el}. Owing to the magnetic nature, the SHO's frequency is directly tunable via external or internal magnetic fields spanning a broad regime from MHz up to GHz frequencies, with the prospect of reaching hundreds of GHz \cite{Bonetti:2009hq,Maehara:2014eo,Naganuma:2014gb,Rippard:2004gya}.

Due to nonlinearities in the magnetic system, an intrinsic connection of the frequency to the oscillation amplitude occurs\cite{Consolo:2010jx}. This allows to tune SHOs directly by the strength of the driving spin current and additionally provides an exceptional wide-ranged synchronization to external signals by frequency entrainment \cite{Wang:2017hl,Slavin:2009fx,Tiberkevich:2014eh}.

However, these nonlinearities also decrease the phase stability of SHOs \cite{Sankey:2005ix,Chen:2016eo,Keller:2009ix}. They promote scattering processes among the different eigenmodes and increase sensitivity to ambient noise\cite{Tiberkevich:2014eh,Sharma:2014ff} . As a consequence, SHOs are limited in their phase stability and output power \cite{Chen:2016eo}. To suppress the nonlinear scattering processes, the active region of the ferromagnetic layer is typically restricted to the nanoscale. This reduces the amount of eigenmodes and results in extended single-mode operation on the expense of absolute output power. Therefore, current research aims at synchronizing either several SHO to each other \cite{Kaka:2005gj,Awad:2016eo} or increasing the active area of SHOs \cite{Luo:2017eb}. 

Recent works show, that the active area of SHOs can be expanded to one-dimensional wires up to the micrometer range \cite{Duan:2014et}. Moreover, it has been demonstrated that slight tapering of the nanowire extends the single-mode operation and reduces their phase noise \cite{Yang:2015hf}. 
Nevertheless, for higher spin current densities, these SHO switch from a single-mode operation to a spectrally broader and polychromatic output.\cite{Yang:2015hf} For this regime, the simultaneous excitation of multiple bullet-modes has been suggested based on electrical measurements and modeled by a nonlinear Ginzburg-Landau equation \cite{Yang:2015hf}, but has not yet been directly observed. Such spin-wave bullets are stabilized by spatiotemporal self-focusing into a two-dimensional dynamically confined soliton. Importantly, in the tapered geometry, the driving torques and demagnetizing fields change continuously along the SHO. This is predicted to result in the stabilization of spin-wave bullets at distinct positions with different frequencies.
In this paper we confirm the proposed formation of two bullets by Brillouin light scattering (BLS) microscopy \cite{Sebastian:2015cka} and measure their localization and frequency. We demonstrate synchronization of the auto-oscillation modes to external signals and investigate the corresponding manipulation of the individual spin-wave bullets. By imprinting an external microwave modulation on the driving current, injection locking of the SHO, amplitude enhancement and linewidth reduction of the individual bullet modes is demonstrated. Moreover, single-mode operation of the SHO is facilitated by suppression of modes different from the selected bullet, while the frequency of the enhanced bullet can be defined by the chosen microwave frequency. In the final step the increase in the amplitude of the spin-wave bullets by an external signal is studied as a function of the microwave power. We find a steep initial increase in auto-oscillation intensity promising for the use of these SHO as microwave detectors or amplifiers.

\section*{Results}

The first section presents the studied SHO geometry and its magnetic configuration. In the next section, the linear spin-wave eigenmodes as a basis of the auto-oscillations are assessed by BLS microscopy measurements and micromagnetic simulations. Subsequently, the formation of spin-wave bullets is confirmed experimentally, providing a description of their characteristics. Finally, injection locking of the SHO and its effect on the individual spin-wave bullets are discussed. All measurements were performed at ambient temperatures.

\begin{figure}[!b]
\centering
\includegraphics[width=16cm]{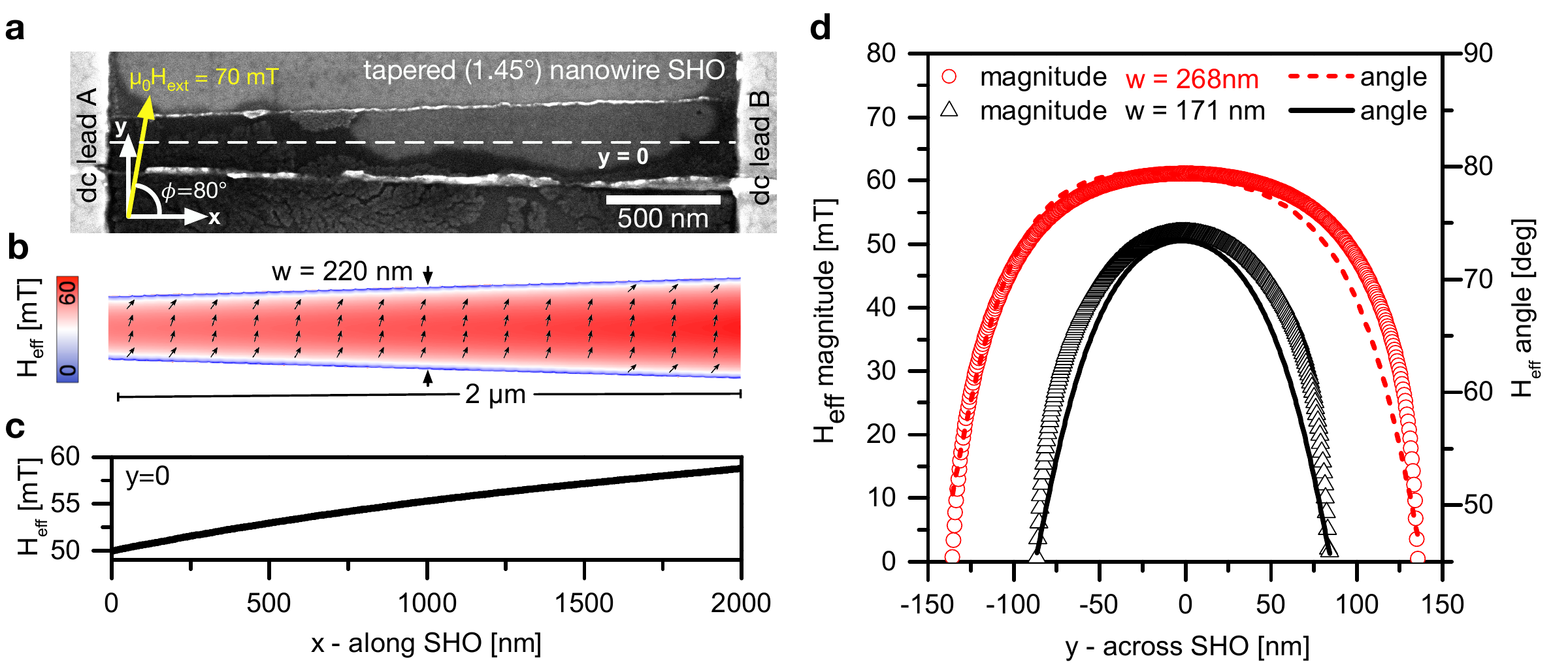}
\caption{Samples. \textbf{a} SEM micrograph of the tapered nanowire SHO. \textbf{b} Micromagnetic simulation of the effective field for an external magnetic field of 70\,mT applied under an angle $\phi = 80^\circ$ with respect to the long axis of the SHO ($x$-axis) . The magnitude is color coded, arrows indicate the orientation of the effective field and magnetization. \textbf{c} Simulated magnitude of the effective field along the SHO ($x$-axis) extracted in the centre of the nanowire ($y=0$). \textbf{d} Simulated effective field distribution across the nanowire (y-axis) for both ends of the active region.}
\label{fig:sample}
\end{figure}

\subsection{Sample description and magnetic configuration.} 

Tapered nanowire spin Hall oscillators (SHO) were fabricated by sputter deposition and electron beam lithography. Following the design of previous studies \cite{Duan:2014et,Yang:2015hf}, the layer-stack consists of Pt(7\,nm)/\ Ni$_{\mathrm{80}}$Fe$_{\mathrm{20}}$(5\,nm)/Al$_{\mathrm{2}}$O$_{\mathrm{3}}$(2\,nm). 
A scanning electron microscopy (SEM) image of the studied nanowire SHO along with the coordinate system is shown in Fig. \ref{fig:sample}\textbf{a}. It has a total length of 6\,\textmu m and is tapered by an angle of 1.45\textdegree . 
Its active region is limited to a length of 1.9\,\textmu m by Au(35\,nm)/Cr(7\,nm) electrodes, that allow to apply $dc$ and $rf$ currents to the SHO. Due to the tapering, the width increases linearly from 171\,nm up to 268\,nm along the active region. More detailed information on the fabrication of the device is given in the method section.
%
%

To set the magnetic state of the nanowire SHO as studied in previous works\cite{Yang:2015hf}, a saturating magnetic field of 300\,mT is applied at an in-plane angle of $\phi = 80^\circ$ with respect to its long axis. The external field is then gradually decreased in strength to 70\,mT. This configuration allows for injection locking of the SHO to a microwave modulation of the bias current as will be discussed later. 
The resulting magnetization and effective field are shown in Fig.~\ref{fig:sample}\textbf{b} as calculated by micromagnetic simulations (see methods). For an external magnetic field applied at $\phi = 80^\circ$, almost perpendicular to the long axis of the nanowire, effective field wells are created at the top and bottom edges of the SHO. These wells and the magnetic orientation are shown in Fig.~\ref{fig:sample}\textbf{d} for the outmost positions at the $dc$ leads. Due to the tapering, the effective field and magnetization changes gradually along the SHO. This spatial dependency of the maximum amplitude of the effective field along the SHO is plotted in Fig.~\ref{fig:sample}\textbf{c}. As expected, the demagnetizing field in the center of the SHO ($y=0$) decreases with increasing width of the nanowire, so that the effective field is found to increase almost linearly along the SHO by about 9 mT.

\begin{figure}[!b]
\centering
\includegraphics[width=16cm]{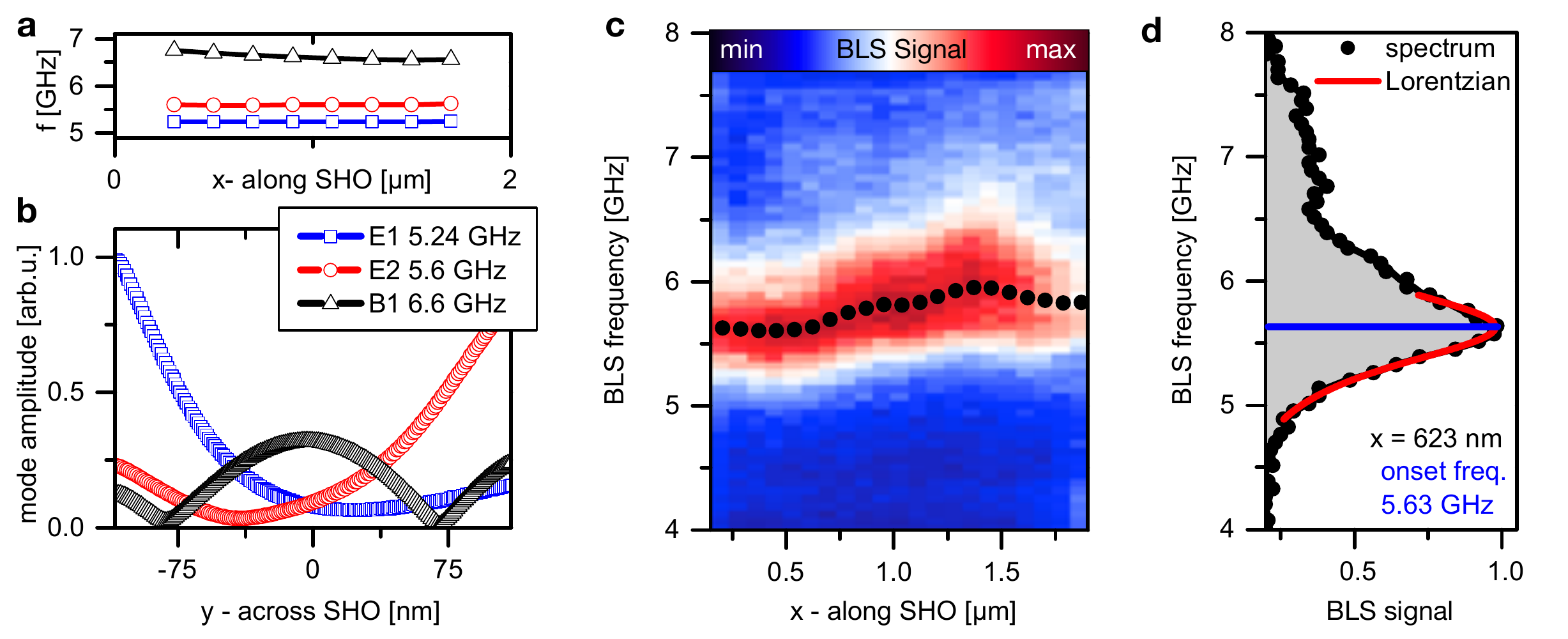}
\caption{Linear spin-wave modes of the SHO. \textbf{a} Mode frequencies along the SHO ($x$-direction) as determined from micromagnetic simulations of the local spin-wave spectrum. \textbf{b} Simulated mode amplitude profile in the center of the SHO (plotted against y-direction) for the three resonances. \textbf{c} Thermal BLS microscopy measurements. Each vertical line represents a spectrum for a fixed position on the SHO with the intensity color coded. The dots indicate the onset frequencies of the spin-wave band that are determined via  Lorentzian fits, as is exemplarily shown in \textbf{d} for the spectrum recorded at $x=623\,\mathrm{nm}$.  }
\label{fig:thermal}
\end{figure}

\subsection{Linear spin-wave dynamics} 

Before addressing the nonlinear auto-oscillations, we investigate the linear spin-wave eigenmodes of the SHO as a basis for its operating characteristics. Therefore, we perform micromagnetic simulations for small excitation amplitudes.
Three spin-wave modes of discrete frequencies and different localization across the SHO (y-direction) are found. Their frequencies and amplitude profiles are plotted in Fig.~\ref{fig:thermal}\textbf{a},\textbf{b}. The two lower frequency modes of 5.24 GHz (E1) and 5.6\,GHz (E2) are localized at the bottom and top edge of the SHO, respectively.  Such modes are well known from transversally magnetized nanowires \cite{Topp:2008bva,Duan:2014jw} and referred to as ``edge'' modes. Since the top and bottom edge of the SHO enclose different angles with the external field, the two edge modes are split in frequency by about 360 MHz. The third linear mode (B1) is localized to the widths center ($y=0$) of the wire (bulk mode) and decreases slightly in frequency along the SHO from $6.75 \textrm{ GHz}$ at the narrow end to $6.55 \textrm{ GHz}$ at the wide end.



We then utilize BLS microscopy \cite{Sebastian:2015cka} to measure thermally excited spin waves at several positions along the SHO experimentally in Fig.~\ref{fig:thermal}\textbf{c}. Each vertical line comprises a spin-wave spectrum with the color code displaying the detected intensity. In each spectrum, the signal maximum closely corresponds to the lowest frequency of the spin-wave band \cite{Cottam:1986bh}. For analysis of this maximum, each spectrum was fitted by a Lorentzian function to the low-frequency side. An exemplary fit is shown in Fig.~\ref{fig:thermal}\textbf{d} for the position at $x=623\,\mathrm{nm}$ yielding an onset frequency of 5.63 GHz. The estimated spin-wave band gap along the SHO is plotted in the BLS intensity graph in Fig.~\ref{fig:thermal}\textbf{c} (black dots). Moving along the SHO within the active region from its narrow to its wide end, it increases by $\sim 300 \textrm{ MHz}$ from about 5.6 to 5.9\,GHz, which we attribute to the increase of the effective field of about 9\textrm{ mT} (see Fig.~\ref{fig:sample}\textbf{c}) combined with an increase in the spin-wave quantization width. 

The experimental observation of one single spin-wave band is in contrast to the three distinct spin-wave resonances found in the simulations. We believe this is attributed to the finite spin-wave coherence length in the real device, which does not permit the simulated quantization into distinct modes. As discussed in the following section, the spin-wave coherence length increases when driving the SHO by a $dc$ current $I_{\textrm{dc}}$, due to the compensation of the dissipative losses. Consequently, we observe discretized spin-wave eigenmodes for this case in the same device.



\begin{figure}[!b]
\centering
\includegraphics[width=16cm]{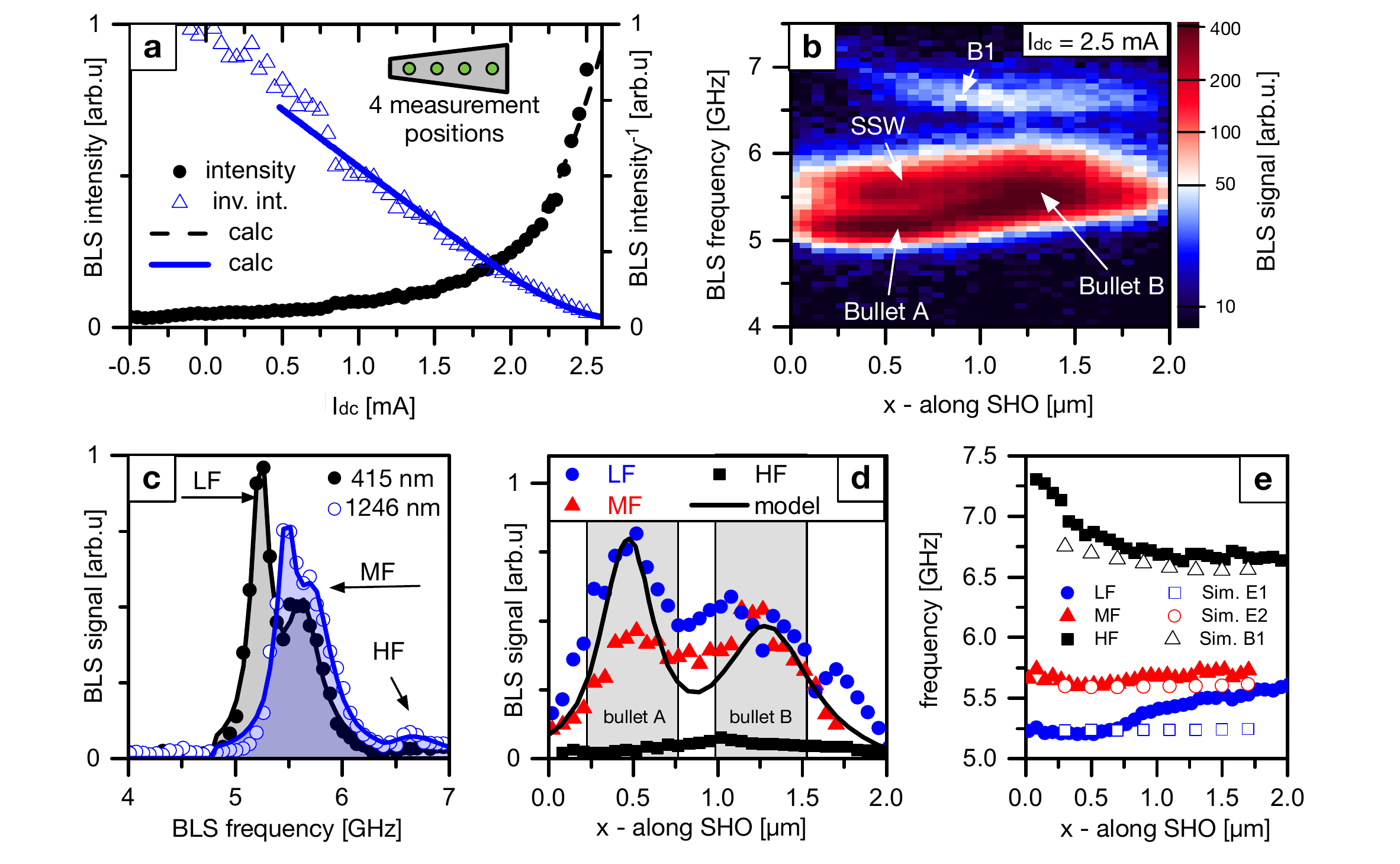}
\caption{Auto-oscillatory modes. \textbf{a} Measured BLS intensity (black dots) integrated over four equidistant $x$-positions and inverse intensity (blue triangles). The output power calculated according to an analytical approximation\cite{Slavin:2009fx} (solid and dashed lines) agrees well with the experimental data when assuming a critical current $\zeta = 2.28 \textrm{ mA}$, nonlinear coefficient $Q=0.8$ and noise $\eta=1$. \textbf{b} BLS spectra recorded for $I_\textrm{dc}=2.5\,\textrm{mA}$ along the SHO with the intensity color coded on a logarithmic scale. Two regions of pronounced intensity and different frequency are observed at $x=480\,\mathrm{ nm}$ and $x=1250\,\mathrm{ nm}$. \textbf{c} Two exemplary BLS spectra measured at different positions for $I_\textrm{dc}=2.5\textrm{ mA}$. A low-frequency (LF), middle-frequency (MF) and high-frequency (HF) peak can be distinguished and fitted by Lorentzian functions. \textbf{d}, \textbf{e} Amplitude distribution and local frequency of the LF, MF, and HF peak along the SHO ($x$-axis). The black line in \textbf{d} shows the amplitude distribution of the auto-oscillating mode according to the proposed formation of two spin-wave bullets\cite{Yang:2015hf}. }
\label{fig:2p5mA}
\end{figure}

\subsection{Bullet formation in the free running SHO} 

To determine the current range for $dc$-driven auto-oscillations, constant currents $I_{\textrm{dc}}$  between $-0.5$ and 2.5\,mA were applied to the nanowire SHO. This regime without any rf modulation on the driving current is referred to as the free running regime of the SHO. The dynamic response was then detected via BLS microscopy at four equidistant positions along the SHO. The black dots in Fig.~\ref{fig:2p5mA}\textbf{a} show the auto-oscillation signal, integrated for all four positions, as a function of $I_{\textrm{dc}}$, while blue triangles represent the inverse of the integrated intensity.  When calculating the output power of the SHO (equation 84b in \cite{Slavin:2009fx}), we obtain best agreement to the measured data assuming a critical current $\zeta= I_C = 2.28\,\textrm{ mA}$, a nonlinear damping coefficient $Q= 0.8$ and a noise background $\eta =1$ (black and blue lines). The critical current $I_C$ corresponds to an averaged current density in the Pt layer of $95\,\textrm{ MA/cm}^2$ and agrees well with typical values for Pt/Ni$_{\mathrm{80}}$Fe$_{\mathrm{20}}$ systems found in literature \cite{Demidov:2011fsb,Demidov:2014ix}. 

For the highest applied current of $I_\textrm{dc}=2.5\,\textrm{ mA}$ we record the spatial characteristic of the auto-oscillations as shown in Fig.~\ref{fig:2p5mA}\textbf{b}. Each vertical line represents a BLS spectrum with the intensity color coded on a logarithmic scale. For this driving current we observe multiple auto-oscillation signals at different frequencies and distinct spatial dependence as predicted for the formation of separate spin-wave bullets.
As discussed in the following, we identify two points of main auto-oscillation intensity as bullet A and bullet B. In addition, two modes are observed for higher frequencies with decreasing signal intensities referred to as ``standing spin-wave'' ,SSW mode, and B1 mode. To assess the validity of this classification into two spin-wave bullets and additional modes B1 and SSW, we analyze the spectra in more detail. As exemplarily presented in Fig.~\ref{fig:2p5mA}\textbf{c}, three auto-oscillation peaks occur simultaneously. For each $x$-position we fit these signal peaks by three separate Lorentzian (solid lines) and distinguish between a low-frequency (LF), middle-frequency (MF) and high-frequency (HF) signal. The extracted spatial dependency of the individual peak amplitudes is presented in Fig.~\ref{fig:2p5mA}\textbf{d}. The corresponding center frequencies are shown together with the simulated spin-wave eigenmodes in Fig.~\ref{fig:2p5mA}\textbf{e}.

Discussing first the auto-oscillation amplitude, the LF and MF peaks contribute strongest to the auto-oscillations. Both of these modes show higher intensities at the positions about $x=500\,\mathrm{ nm}$ and $x=1250\,\mathrm{ nm}$. The localization to these two distinct positions indicates the formation of two bullet modes. The calculation of this state according to the nonlinear Ginzburg-Landau model\cite{Yang:2015hf} is shown as a black line in Fig.~\ref{fig:2p5mA}\textbf{d} for a given moment in time. As a guide to the eye, the modeled bullet positions are schematically indicated by grey boxes in the background of the data. 
We obtain best agreement between the calculated amplitude distribution and its smaller value for the bullet mode localized in the wider part of the wire at $x=1250\,\mathrm{ nm}$ with the experimental data for the LF signal. However, for the experimentally recorded amplitude of the LF signal the extrema are less pronounced, which very likely results from the convolution of the measured intensities with the finite spatial resolution of the BLS microscope of about $340\textrm{ nm}$.
Extracting the frequency of the LF signal at these two positions (Fig.~\ref{fig:2p5mA}\textbf{e}) yields bullet frequencies of 5.2 and 5.45 GHz, respectively. Hence, the spectral separation due to the tapered geometry is determined to be 250 MHz. This is in good agreement with the modeled value of 190 MHz\cite{Yang:2015hf}. Therefore, we identify two spin-wave bullets corresponding to the LF signal.

In addition to these two bullet modes, a standing spin wave, the SSW mode, that is quantized along the length of the SHO ($x$-direction) is observed corresponding to the MF signal. The assumption of a standing spin wave is substantiated by the constant mode frequency of about 5.6 GHz along the entire SHO and high degree of spatial symmetry of its amplitude. This mode is also in accordance with an expected increase in the coherence length by the spin-transfer torque, promoting quantization effects. Therefore, we conclude that this auto-oscillation originates from a linear spin-wave eigenmode quantized along the active region of the SHO by the abrupt change in effective field and damping at the $dc$ leads. Qualitative comparison of the mode frequency with the simulated spin-wave eigenmodes indicates, that this standing wave is likely localized to the bottom or top edge of the SHO (E1 or E2). 

The third HF signal is only about 10\% in amplitude compared to the aforementioned signals. The corresponding mode exhibits a higher frequency and similar frequency dependence along the SHO as the bulk mode B1 that is observed in the simulations. Therefore, we attribute this mode to the linear spin-wave mode localized across the SHO nanowires center ($y=0$). Due to its low intensity, this mode is expected to be negligible for our discussion of the auto-oscillations in the tapered SHO. In the following, we focus on the manipulation of the dominant spin-wave bullets and SSW mode during synchronization of the SHO by injection locking.

\subsection{Injection locking of the SHO} 

We now discuss synchronization of the SHO to an external signal in form of a well defined microwave modulation of the $dc$ current. For nano-contact SHOs, this approach has been shown to lead to frequency entrainment and an increase in coherence within a certain frequency range, which we here refer to as the \textit{locking window}\cite{Rippard:2013bb,Lebrun:2015ei,Ulrichs:2014gka,Tsunegi:2016ka}. In this work, we investigate this synchronization for the micrometer long tapered nanowire SHO. These experiments demonstrate the SHOs general ability to synchronize to other oscillators or external sources, but are particularly interesting for maintaining its single-mode operation as well as improving its spectral purity, despite the multitude of available eigenmodes. In addition, the bullets amplitude, localization and frequency can be manipulated via the injection of external signals.

For the experiment, we combine a microwave signal with a constant current source via a common bias Tee. This allows the simultaneous application of both an $ac$ current $I_\textrm{ac}$ and a $dc$ current $I_\textrm{dc}$ to the SHO. To exclude pronounced excitation of spin waves by the microwave current itself, we measure the spin-wave spectra for a microwave power of nominally 0.06\,mW for the inactive SHO, $I_\textrm{dc} =0\,\mathrm{ mA}$. As expected, only a slight increase in the detected intensities well below 10\% is observed compared to the thermal spectra. 
Hence, direct excitation of magnetization dynamics by the $ac$ currents can be neglected for these small microwave powers.

\begin{figure}[!b]
\centering
\includegraphics[width=16cm]{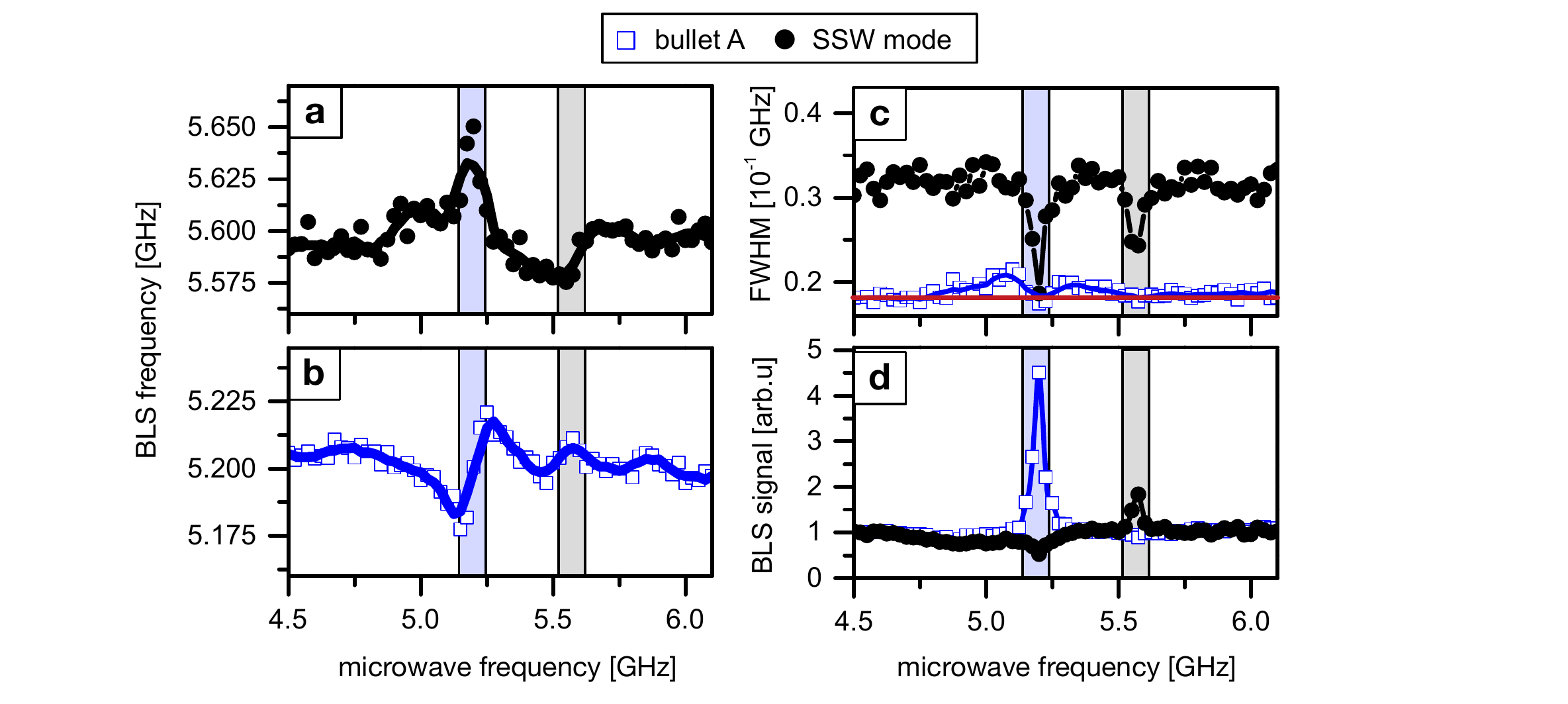}
\caption{Injection locking of the SHO. Microwave frequencies are applied in the range between 4.5 and 6.2\,GHz at a nominal {\it ac} output power of 0.06\,mW and a driving current of $I_{dc}=2.5 \textrm{ mA}$. Measurement position on bullet A at $x=500\,\mathrm{nm}$. The frequencies ({\textbf a} and {\textbf b}), amplitudes ({\textbf c}) and FWHM ({\textbf d}) are shown for bullet A (blue hollow squares) and the SSW mode (black dots). The black and blue solid lines serve as a guide to the eye. The estimated locking intervals of about 100\,MHz width centered around the auto-oscillation frequencies of 5.2 and 5.57\,GHz are marked by light-blue and grey boxes, respectively. Within these windows, frequency locking, amplitude enhancement/suppression, and linewidth reduction (detection limit marked by a red line in \textbf{c}) are observed.}
\label{fig:locking1}
\end{figure}

Subsequently, the injection locking of the SHO, driven at $I_\textrm{dc} =2.5\,\mathrm{mA}$, to this small amplitude $rf$ signal is analyzed by recording the auto-oscillations at the position of bullet A ($x=500\,\mathrm{nm}$). In the experiments, the microwave frequency is swept over the spectral range of auto-oscillations covering the bullet and SSW mode.  In order to determine their frequency, amplitude and linewidth, each spectrum is fitted by two Lorentzian. The obtained data is summarized in Fig.~\ref{fig:locking1} presenting the extracted frequencies (\textbf{a} and \textbf{b}), linewidths (\textbf{c}) and amplitudes (\textbf{d}) for the bullet mode (blue squares) and the SSW mode (black dots).

We observe synchronization of the SHO to the external RF signal within two separate locking windows centered around the free-running auto-oscillation frequencies of 5.2 and 5.57\,GHz. This is seen by the frequency entrainment of the modes and the external signal within the locking windows highlighted by light-blue and grey boxes in the background of the data in Fig.~\ref{fig:locking1}\textbf{a,b}.
These approximately 100 MHz-wide locking windows are estimated by the points of inflection in the measured frequency dependencies as well as simultaneous linewidth reduction presented in Fig.~\ref{fig:locking1}\textbf{c}. The free-running linewidth of the bullet mode has been shown to be smaller than 100 MHz by electrical measurements\cite{Yang:2015hf}. However, our recorded values are limited by the spectral resolution of the experimental set up of about 190 MHz (red line in Fig.~\ref{fig:locking1}\textbf{c}). Nevertheless, the Full width at half maximum (FWHM) linewidth of the SSW mode is found to be on the order of 300 MHz and reduces significantly within the locking windows. This reduction clearly shows the improved coherence and stabilization of the auto-oscillations by the external signal. 
Additionally, it may also be attributed to reduced mode hopping and smaller spatial fluctuations of the bullet modes for the injection-locked SHO.

Analyzing the mode amplitudes in the locked regime (Fig.~\ref{fig:locking1}\textbf{d}), reveals a pronounced amplification of the locked mode, whereas the other mode is simultaneously suppressed in amplitude. This suppression excludes a sole enhancement of the auto-oscillations by direct excitation via the microwave signal and substantiates a redirection of energy towards the single-mode operation in the synchronized state.

\begin{figure}[!b]
\centering
\includegraphics[width=16cm]{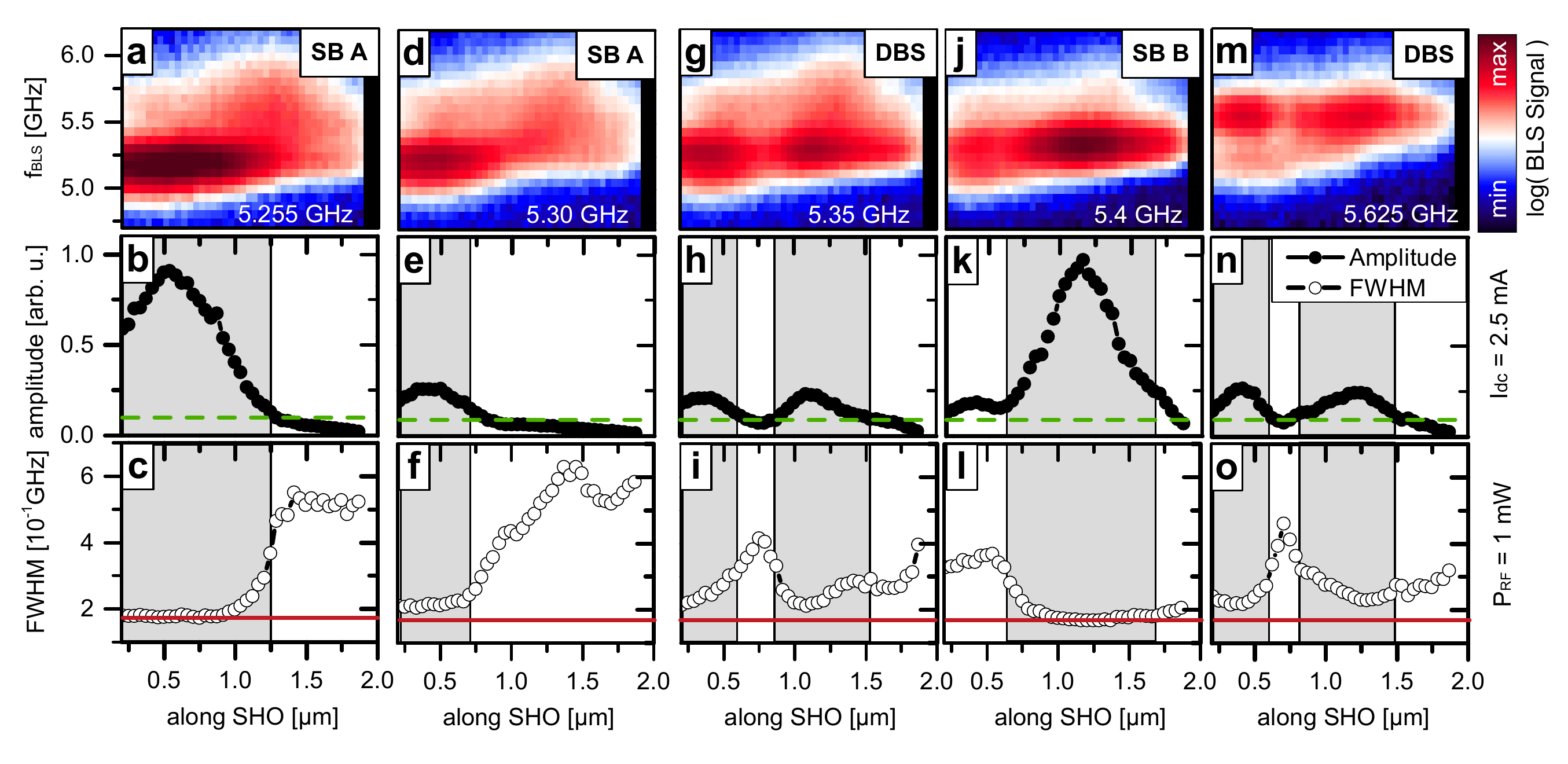}
\caption{ Spatial distribution of amplitude and linewidth during injection locking. Microwave frequencies of 5.255\,GHz ({\textbf a}-{\textbf c}), 5.3\,GHz ({\textbf d}-{\textbf f}), 5.35\,GHz ({\textbf g}-{\textbf i}), 5.4\,GHz ({\textbf j}-{\textbf l}) and 6.2\,GHz ({\textbf k}-{\textbf m}) are applied with a nominal  {\it ac} power of 1\,mW and a driving current of $I_\textrm{dc} = 2.5 \textrm{mA}$. ({\textbf a},{\textbf d},{\textbf g},{\textbf j},{\textbf m}) Spatially resolved BLS measurements. ({\textbf b},{\textbf e},{\textbf h},{\textbf k},{\textbf n}) local amplitude of the auto-oscillation signal  and its FWHM ({\textbf c},{\textbf f},{\textbf i},{\textbf l},{\textbf o}) as determined from Lorentzian fitting of each spectra. For different frequencies of the injected microwave signal, locking of either only one bullet (SB A for locked bullet A, SB B for locked bullet B) or two spatially separated regions (DBS) is observed. In the locked regions, the linewidth is reduced accompanied by an increase in oscillation amplitude. In addition, the maximum amplitude of the free-running auto-oscillation (green dashed line) and spectral resolution of the BLS microscope (red line) is indicated.}
\label{fig:lockedmodes}
\end{figure}

For deeper analysis, we now discuss the spatial and spectral characteristic of the locked auto-oscillations, which we image via spatially-resolved BLS measurements along the SHO for different frequencies of the injected microwave current. Figure~\ref{fig:lockedmodes} summarizes the results for microwave frequencies of 5.255\,GHz (\textbf{a}-\textbf{c}), 5.3\,GHz (\textbf{d}-\textbf{f}) 5.35\,GHz (\textbf{g}-\textbf{i}), 5.4\,GHz (\textbf{j}-\textbf{l}), and 5.625\,GHz (\textbf{m}-\textbf{o}). All measurements were performed at an increased nominal {\it ac} output power of $1$\,mW and a direct current of $I_\textrm{dc} =2.5\,\mathrm{mA}$ to enhance the frequency entrainment and amplification. They are normalized to an identical optical reference signal and shown on the same logarithmic color scale to achieve comparable amplitudes. Grey boxes in the background of the data indicate the synchronized areas as derived from the reduction in linewidth and enhanced amplitude. We observe different regimes in which either only bullet A or B (single bullet state ``SB A'' and ``SB B'') or both bullet modes (double bullet state ``DBS'') couple to the microwave signal. 



We first focus on the synchronization of one single individual bullet, when applying microwave frequencies of 5.255 or 5.4\,GHz. These two frequencies are close to the free-running frequencies of bullet A and B, which oscillate at 5.2 and 5.45 GHz, respectively. The amplitude analysis in Fig.~\ref{fig:lockedmodes}\textbf{b},\textbf{n} show the amplification of the individual bullets. As a guide to the eye, the maximum amplitude of the free-running auto-oscillation outside of the locking regime is indicated by green dashed lines. Both bullets remain at their initial position around $x=500\,\mathrm{nm}$ (Fig.~\ref{fig:lockedmodes}\textbf{a}) and $x=1250\,\mathrm{nm}$ (Fig.~\ref{fig:lockedmodes}\textbf{j}), but exhibit amplitudes that are by a factor of 8 to 10 larger compared to the free running values and a linewidth reduced to the detection limit (Fig.~\ref{fig:lockedmodes}\textbf{c},\textbf{l}). 
When the microwave frequency is slightly increased from 5.255 GHz to 5.3 GHz the amplitude of bullet A reduces and the spatial extent of the locked-regime decreases accompanied by a slight shift in the bullet position to about $x=400\textrm{ nm}$. Further incrementing the external frequency by another 50 MHz to 5.35\,GHz leads to a transition in the locking characteristic. The auto-oscillatory state clearly shows two maxima with equal intensities at distinct positions along the nanowire (Fig.~\ref{fig:lockedmodes}\textbf{g}-\textbf{i}). We attribute this observation to a partial locking of both bullets to the external signal with increased hopping between the aforementioned two single-bullet operations. This is substantiated by the increased linewidth and reduced amplitudes compared to the single-bullet states.

Surprisingly, for an external frequency of 5.625\,GHz, higher than the one of the bullets, a state of similar amplitude and linewidth dependence is observed (Fig.~\ref{fig:lockedmodes}\textbf{m}-\textbf{o}). On first glance, this could be attributed to synchronization of the SSW mode spanning along the SHO at a free-running frequency about 5.6 GHz, very close to the external one. 
However, the linewidth of a standing spin-wave resonance is expected to stay constant spatially. In contrast to this, we observe an increase in linewidth at about $x=700\textrm{ nm}$ between the amplitude maxima. This observations seems to indicate partial locking of the individual bullets. We do not discuss the nature of this state in this work in greater detail, but focus on the amplification of bullet A with increasing microwave power.
\begin{figure}[tb]
\centering
\includegraphics[width=6cm]{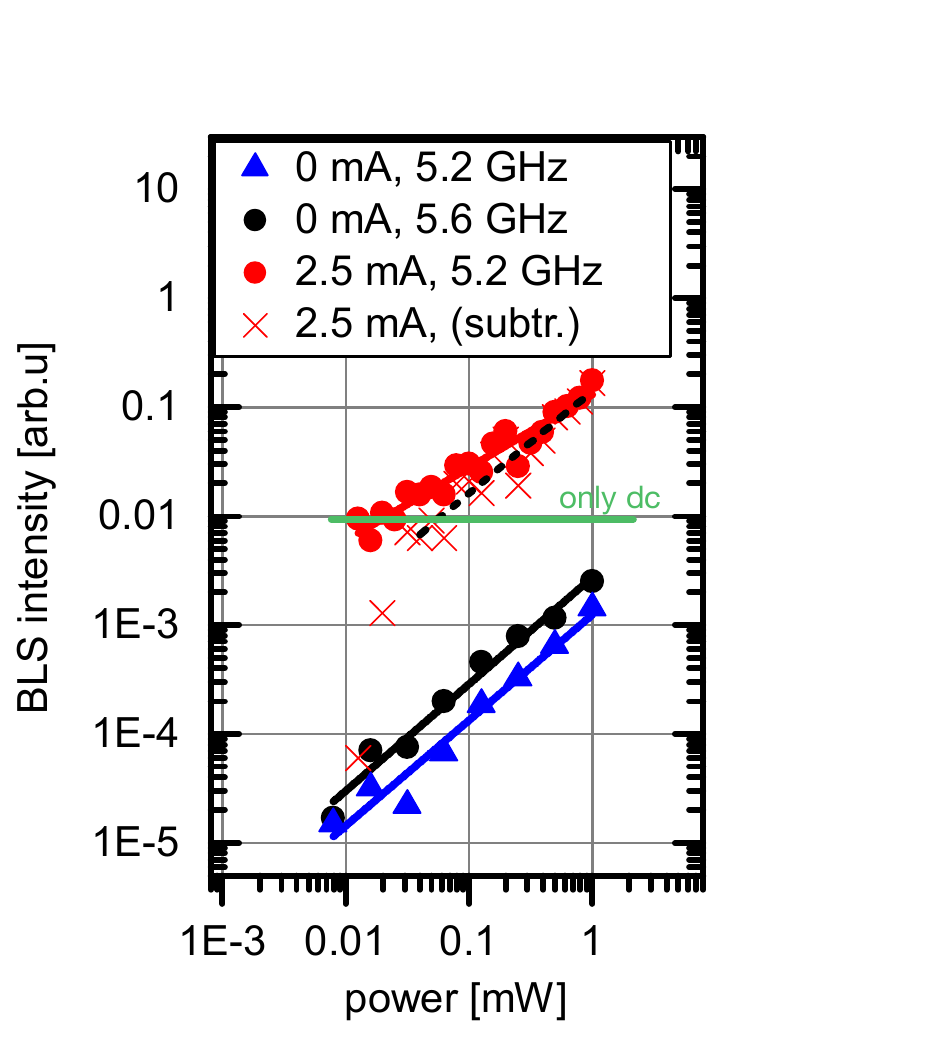}
\caption{Auto-oscillation intensity as a function of the injected microwave power in the locked regime for $I_{\textrm{dc}}=2.5\textrm{ mA}$. The signal amplitude measured on the position of bullet A for $x=500\textrm{ nm}$ and and a microwave frequency of $5.2\textrm{ GHz}$ is shown as red dots. Linear fitting of the data suggests a power law with an exponent of 0.67 (red line). For comparison the recorded BLS intensity without an externally applied signal is shown as a green horizontal line. The signal amplitude subtracted by this base-intensity is additionally shown as red crosses. Blue triangles and black squares give the amplitude of the directly excited magnetization dynamic by the RF current when the SHO is turned off ($I_{dc}=0$) for microwave frequencies of 5.2 GHz and 5.6 GHz, respectively.}
\label{fig:powerdep}
\end{figure}

The measurement position is fixed again on the bullet A, $x=500\textrm{ nm}$, and different external signal strengths, that cover a power range between 0.008 mW and 1\,mW at a frequency of 5.2\,GHz, are applied to the SHO for $I_\textrm{dc}=2.5\,\mathrm{mA}$.
Figure~\ref{fig:powerdep} presents the recorded intensities of bullet A (red dots) as a function of the microwave power. A linear fit of the amplified signal (red line) results in a power exponent of 0.67 close to $2/3$. To discuss the enhancement of the auto-oscillations by injection locking, the green line shows the measured intensity when only applying the $dc$ current in the free running regime. As expected, this base intensity is identical to the signal strength for very low microwave powers within the noise of the measurements. The difference of the intensity during injection locking (red dots) and the base intensity (green line) is plotted as red crosses in the same graph. According to analytical calculations\cite{Slavin:2005ga}, this difference is expected to follow a square root dependence on the microwave power. Interestingly, the data show a much steeper increase, in particular for microwave powers below 0.04 mW. 

We believe this can be attributed to thermal effects on the phase stability of the auto-oscillations. In the abscence of an external signal, the ambient noise decreases the coherence of the auto-oscillations resulting in increased linewidth and reduced output power. When increasing the microwave signals above the thermal noise floor, the phase is defined by the external signal leading to injection locking. This results in a steep increase in the auto-oscillation signal with increasing microwave power due to the higher degree of coherence.  
For microwave powers above 0.04 mW, a less pronounced linear increase in amplification with applied power of the microwave signal is observed as indicated by the linear fit of slope $0.95\pm0.08$ (dashed black line) as guide to the eye. This indicates the transition to simple additional direct excitation of spin-wave dynamics with further increasing the microwave power, which is connected to an increase in the precession cone angle, after phase coherence is established.

To estimate the influence of the direct excitation of dynamics by the microwave current, we additionally perform measurements when the SHO is turned off ($I_\textrm{dc}=0\,\mathrm{mA}$) and detect the intensities for the frequency of bullet A (blue triangles), as well as the frequency of maximum excitation efficiency of 5.6 GHz (black dots). As can be clearly seen, the strength of the directly excited dynamics is well below the amplification values and follows a linear power dependence confirmed by the slopes of $0.98\pm0.06$ and $0.97\pm0.08$. Hence, we conclude that two regimes of amplification of the auto-oscillations are observed. Up to microwave powers of about 0.04 mW the coherence is improved by overcoming the fluctuations stemming from the thermal noise floor via injection locking of the SHO to the external microwave signal. For higher microwave powers further amplification of the auto-oscillation signal is observed with a less pronounced increase. This is promising for utilizing such SHO as sensitive microwave detectors. 
 
%
%

\section*{Discussion and conclusion}
 
In this work, the auto-oscillations in SHOs of tapered nanowire geometry are studied experimentally. 
Spatially resolved BLS microscopy is used to decompose the oscillations into their spectral contributions and reveal their individual localization. 
Our experimental study confirms that higher driving currents excite two separate auto-oscillatory modes, which oscillate at different frequencies and are localized at separate positions spaced by about 750 nm.
These modes are identified as nonlinear self-localized spin-wave bullets, stabilized by the tapered geometry, in agreement with previous electrical characterization and a nonlinear Ginzburg-Landau model \cite{Yang:2015hf}. 
%
We control the state of auto-oscillation utilizing externally injected microwave signals with various frequencies and signal powers.

In particular, it is demonstrated that the auto-oscillations can be tuned by the $rf$ signal so that either one or both bullets are synchronized to the external source and enhanced in amplitude. Simultaneously, all other modes are suppressed promoting single mode operation. This allows the manipulation of the tapered SHO simply by small changes in the injected microwave frequency or power. Within the entire investigated regime of externally imprinted microwave powers up to 1\,mW, signal amplification is observed with an initial steep increase in output power attributed to overcoming the thermal noise and increase in coherence of the oscillations.
This grants an additional possibility of configuring the SHO output with the prospect of utilization in microwave amplifiers or detectors. As an important advantage over straight nanowire geometries, the tapered nanowire design offers reduced phase noise, extended single-mode operation and is promising for the realization of several spin-wave bullets with distinct frequencies and position when increasing the oscillator length. Their coherence, mutual interaction and individual reaction to external signals can be targeted in future time-resolved studies in these SHO. 
In particular, combining various bullets with distinct frequencies and localization in one device is promising for the realization of wide locking windows and a high degree of tunabilty.




\section*{Methods}

\subsubsection*{BLS microscopy}
The spin-wave intensity was measured locally by means of Brillouin light scattering microscopy (BLS microscopy) at room temperature\cite{Sebastian:2015cka}. Light from a continuous-wave, single-frequency 532\,nm solid-state laser is focused to a laser spot and scanned over the sample-surface. The back-scattered light is then analyzed using a six-pass Fabry-Perot interferometer TFP-2 (JRS Scientific Instruments), where the BLS intensity is proportional to the square of the amplitude of the magnetization dynamics at the laser spot position. For the conditions of the presented measurements the spectral and spatial resolution was experimentally determined to be $190 \pm 30\,\textrm{MHz}$ and $340\pm 50\,\textrm{nm}$, respectively (Full Width Half Maximum values). 

\subsubsection*{Sample fabrication} \label{sec-samplefab}

Thin films are grown by magnetron sputter deposition. A 5 nm thick Pt film is deposited on an Al$_2$O$_3$(0001) substrate at 585 \textdegree C and subsequently annealed in-situ for 1 hour at the same temperature. The tapered nanowire SHO is defined on the 5 nm Pt film by means of e-beam lithography, sputter deposition, and lift-off. For the sputter deposition, the sample is briefly treated to an Ar plasma cleaning etch and then the remaining stack structure Pt(2 nm)/Ni$_{80}$Fe$_{20}$(5 nm)/Al(2 nm) is immediately deposited in-situ at room temperature. The Al(2 nm) capping layer is naturally oxidized and serves to protect the Ni$_{80}$Fe$_{20}$ layer from oxidation. The Cr(7 nm)/Au(35 nm) leads are defined by means of e-beam lithography, e-beam evaporation, and subsequent lift-off. Finally, Ar plasma etching is used to remove the remaining exposed 5 nm Pt layer (everywhere but under the tapered nanowire and leads).


\subsubsection*{Micromagnetic simulations}
Micromagnetic simulations are performed using the software package MuMax3 \cite{Vansteenkiste:2011bx}. A  2.9 \textmu m long tapered magnetic wire of 220\,nm average width, 5\,nm thickness and a tapering angle of 1.45\textdegree~is modeled. The geometry is then represented on a 1024 x 256 x 1 grid, resulting in approximate cell sizes of  2.9 x 1.2 x 5 nm$^3$, below the exchange length of Py. The saturation magnetization $M_S = 530\,\textrm{kA/m}$ and exchange stiffness $A=5\,\textrm{ pJ/m}$  are set according to the experimentally determined values for this type of nanowire \cite{Duan:2014et}. To excite and study the spin-wave dynamics, a homogeneous out-of-plane pulse with a time dependency of $\textrm{sinc}($40$ \textrm{ GHz} \cdot 2\pi \cdot t)$ and a damping parameter $\alpha_\textrm{G} = 0.007$ are used. The magnetic dynamics is saved every 20 ps over a duration of 25\,ns in the simulation. Subsequently, the time trajectory of the magnetization of each cell is Fourier transformed to yield its power spectrum. The mode profiles are then calculated by windowed Fourier backtransformation for the mode frequency.  

%
%


\bibliography{bibliography}


\section*{Acknowledgements}
The authors acknowledge financial support from the Deutsche Forschungsgemeinschaft within programme SCHU 2922/1-1. The work at UCI was supported by the National Science Foundation through Awards No. DMR-1610146, No. EFMA-1641989 and No. ECCS-1708885. We also acknowledge support by the Army Research Office through Award No. W911NF-16-1-0472 and Defense Threat Reduction Agency through Award No. HDTRA1-16-1-0025. K.S. acknowledges funding within the Helmholtz Postdoc Programme. K.W. thanks A. K\'akay for discussion on the experimental results and providing computational infrastructure together with T. Schneider.

\section*{Author contributions}

A.S. and L.Y. fabricated the samples. K.W. and T.H. conducted the experiment(s) and analyzed the experimental data. All authors interpreted and discussed the results. K.W., K.S. and H.S. co-wrote the manuscript. All authors discussed and reviewed the manuscript.


\section*{Additional information}

\textbf{Competing Interests:} The authors declare no competing interests.

\end{document}